\documentclass[prl,showpacs,twocolumn,superscriptaddress]{revtex4-1}
\usepackage{amssymb}
\usepackage{amsmath}
\usepackage{graphicx}

\newcommand{\ph}[1]{\psi^{\vphantom{*}}_{#1}}
\newcommand{\pd}[1]{\psi^*_{#1}}

\begin{document}

\title{Pseudo diamagnetism of four component exciton condensates}
\author{Yuri G. Rubo}
 \affiliation{Centro de Investigaci\'on en Energ\'{\i}a, Universidad Nacional
 Aut\'onoma de M\'exico, Temixco, Morelos, 62580, Mexico}
\author{A. V. Kavokin}
 \affiliation{Groupe d'Etude des Semi-conducteurs, UMR 5650,
 CNRS--Universit\'{e} Montpellier 2, \\
 Place Eug\`{e}ne Bataillon, 34095 Montpellier Cedex, France}
 \affiliation{School of Physics and Astronomy, University of Southampton,
 Highfield, Southampton SO17 1BJ, UK}

\date{March 23, 2011}

\begin{abstract}
We analyze the spin structure of the ground state of four-component exciton
condensates in coupled quantum wells as a function of spin-dependent
interactions and applied magnetic field. The four components correspond to the
degenerate exciton states characterized by $\pm2$ and $\pm1$ spin projections
to the axis of the structure. We show that in a wide range of parameters, the
chemical potential of the system increases as a function of magnetic field,
which manifests a pseudo-diamagnetism of the system. The transitions to
polarized two- and one-component condensates can be of the first-order in this
case. The predicted effects are caused by energy conserving mixing of $\pm2$
and $\pm1$ excitons.
\end{abstract}

\maketitle


\emph{Introduction.}---Cold exciton gases formed by indirect excitons in
coupled quantum wells (CQW) represent a solid state bosonic system very rich in
new quantum coherent phenomena \cite{Butov02}. Indirect excitons are
semiconductor crystal excitations with unique properties: they have long
lifetime and spin-relaxation time, can cool down to the temperatures well below
the temperature of quantum degeneracy, can travel over large distances before
recombination, and can be \textit{in situ} controlled by applied voltage
\cite{Butov04,Rapaport04,Chen05,Haque06,Yang10}. The indirect excitons form a
model system both for the studies of fundamental properties of light and matter
and for the development of conceptually new optoelectronic devices
\cite{Kuznetsova10}. Being formed by heavy holes and electrons, the indirect
excitons may have four allowed spin projections to the axis normal to quantum
well plane: $-2$, $-1$, $+1$, $+2$ \cite{Leonard09}. The excitons with spin
projections $-1$ and $+1$ can be coupled to light. Recombining, they emit left-
and right-circularly polarized photons, respectively. The exciton states with
spin projections $-2$ and $+2$ correspond to the optical transitions forbidden
in the dipole approximation. These states are usually referred to as "dark
excitons". In the absence of external magnetic field dark and bright states of
indirect excitons are nearly degenerate. Thus, the gas of indirect excitons can
be treated as a four-component degenerate Bose gas. Very interestingly, two
components of this gas can be directly studied by the photoluminescence
spectroscopy, while two other components are hidden from the observer.
Moreover, due to the interplay between dipole-dipole repulsion of excitons and
spin dependent exchange interactions of particles, the interactions of indirect
excitons are strongly sensitive to their spin and polarization. These features
make the gases of indirect excitons a unique system whose thermodynamical
properties are expected to be quite different from the properties of
multicomponent superfluids \cite{Salomaa87} and atomic condensates
\cite{Ueda10} studied before. In this Letter we analyze theoretically the spin
and polarization state of Bose-Einstein condensates of indirect excitons (XBEC)
at zero temperature. We account for all exciton interaction terms allowed by
symmetry and study how the polarization of XBEC changes as a function of the
interaction constants and applied magnetic field. We predict the diamagnetic
shift of the chemical potential of the exciton condensate if a specific
condition on interaction constants is fulfilled.

\emph{Condensate at zero magnetic field}---We characterize XBEC by a
four component order parameter $\psi_{m}$ with $m=\pm 1,\pm 2$. The
indices $m$ denote four allowed exciton spin projections on the
structure axis. The Hamiltonian density $H$ in the absence of applied
magnetic field can be written as ($\hbar=1$)
\begin{multline}
 \label{Ham0}
 H=\frac{1}{2M}\sum_m|\nabla\psi_m|^2-\mu n \\
 +\frac{1}{2}\sum_{m,m^\prime}V_{m,m^\prime}|\psi_m|^2|\psi_{m^\prime}|^2
 \\
 +W(\pd{+2}\pd{-2}\ph{+1}\ph{-1}+\pd{+1}\pd{-1}\ph{+2}\ph{-2}).
\end{multline}
Here $n=\sum_m|\psi_m|^{2}$ is the total polariton concentration, $M$ is the
exciton mass, and $V_{m,m^\prime}=V_{m^\prime,m}$. In a general case,
exciton-exciton interactions are described by five independent constants which
come from the amplitude of spin-independent electron-hole interaction and
amplitudes of electron-electron and hole-hole interactions in the triplet
(parallel spins) and singlet (antiparallel spins) configurations. In the
following, we denote $V_{m,m}=V_0>0$, which describes interactions between
particles having identical spins and accounts for the dipole-dipole and exchange
repulsion. The other parameters are $V_{+2,+1}=V_{-2,-1}=V_0-V_e$,
$V_{+2,-1}=V_{-2,+1}=V_0-V_h$, and $V_{+2,-2}=V_{+1,-1}=V_0-V_x$, where
$V_{e,h,x}$ describe exchange interactions between the particles with opposite
spins: electrons, holes, and excitons, respectively. In realistic CQW the
dipole-dipole repulsion dominates over the exchange interaction constants, but
the polarization state of exciton condensate is governed by small
spin-dependent interaction parameters $W$ and $V_{x,e,h}$.

The last term in (\ref{Ham0}) describes the mixing between dark and bright
excitons. It appears due to possibility of scattering of two bright excitons
into two dark ones and vice versa \cite{Ciuti98}. This term is switched off if
one or more exciton spin components become empty. On the other hand, if all
components are occupied, this term reduces the energy of the condensate
independently of the sign of $W$. This is assured by minimization of $H$ over
the phase factor $(\theta_{+2}+\theta_{-2}-\theta_{+1}-\theta_{-1})$, where
$\theta_m$ is the phase of the $m$-th component of the order parameter:
$\psi_m=A_me^{i\theta_m}$. In the following we set $W>0$.

The ground state of XBEC is defined by the signs of parameters
\begin{subequations}
 \label{uPar}
\begin{eqnarray}
 u_0=n(W+V_x+V_e+V_h)/4, \\
 u_x=n(W-V_x+V_e+V_h)/4, \\
 u_e=n(W+V_x-V_e+V_h)/4, \\
 u_h=n(W+V_x+V_e-V_h)/4,
\end{eqnarray}
\end{subequations}
as it is shown in Fig.~1, where we denoted
$u_\mathrm{min}=\min\{u_x,u_e,u_h\}$. Note that it is $u_\mathrm{min}$
that defines which particular two-component condensate (TCC) is realized. For
example, if $V_{x}>W+V_{e}+V_{h}$ and $u_\mathrm{min}=u_{x}<0$, either
pure dark or pure bright condensate is formed. If $V_{e}>W+V_{x}+V_{h}$ and
$u_\mathrm{min}=u_{e}<0$, the dark and bright components become circular with
the same sign of circular polarization. Finally, for $u_\mathrm{min}=u_{h}<0$,
the dark and bright components are also circular, but of the opposite signs.

\begin{figure}[b]
\includegraphics[width=2.75in]{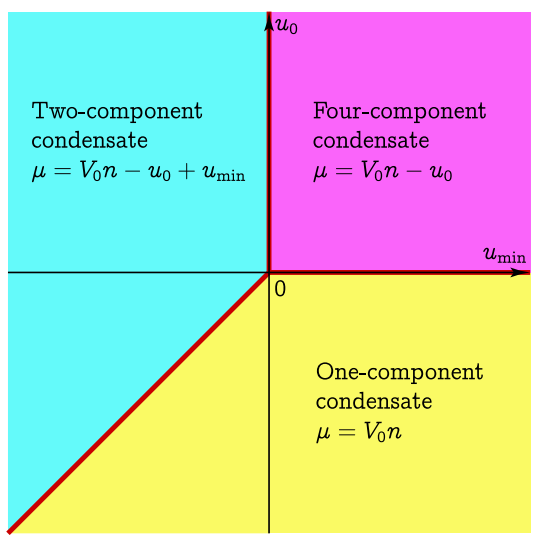}
 \caption{Showing the state of the exciton condensate for different values of
 interaction parameters defined in the text.}
\end{figure}

In what follows, we consider the most interesting and presumably most
experimentally relevant case of a four-component condensate (FCC), where all
parameters (\ref{uPar}) are positive. The chemical potential in this case is
\begin{equation}
 \label{mu0}
 \mu=\mu_{0}=V_{0}n-u_0.
\end{equation}

The excitation spectrum of four-component condensate can be found in the usual
way \cite{LPStatPhys2} by linearizing the Gross-Pitaevskii equation generated
by Hamiltonian (\ref{Ham0}) with respect to small plane-wave perturbation. The
spectrum consists in four branches. Three branches have the Bogoliubov dispersion,
$\varepsilon=\sqrt{\varepsilon_0(k)[2u+\varepsilon_0(k)]}$, where
$\varepsilon_0(k)=k^2/2M$ and $u$ takes the values of $\mu_0$, $u_e$, and
$u_h$. The fourth branch is gaped in $k=0$ and the quasiparticle energy is
\begin{equation}
 \label{GapedQP}
 \varepsilon_x(k)=\sqrt{[Wn+\varepsilon_0(k)][2u_x+\varepsilon_0(k)]}.
\end{equation}
The gap $\Delta=\sqrt{2Wn u_x}$ appears due to the mixing term of Hamiltonian
(\ref{Ham0}). We note that its physical origin is the same as of the gap in BCS
superconductors: namely, the \emph{pare scattering}.

\emph{Effects of Zeeman splitting}---Weak applied magnetic fields affect the
exciton condensate mainly due to addition of the Zeeman splitting term $H_{Z}$
to the Hamiltonian,
\begin{equation}
 \label{HamZ}
 H_Z=-\frac{1}{2}(\omega_1s_1+\omega_2s_2),
\end{equation}
where $\omega _{1,2}=g_{1,2}\mu_{B}B$ are the Zeeman splitting energies for
single excitons and we introduced the pseudospins of the components $%
s_{1}=|\psi _{+1}|^{2}-|\psi _{-1}|^{2}$ and $s_{2}=|\psi _{+2}|^{2}-|\psi
_{-2}|^{2}$. The $g$-factors of dark and bright excitons are $%
g_{1}=g_{h}-g_{e}$ and $g_{2}=g_{h}+g_{e}$, where $g_{e}$ and $g_{h}$
are $g$-factors of electrons and heavy-holes, respectively. Both the
values and the signs of $g$-factors can be different as they depend
substantially on the quantum well widths \cite{Snelling92,Bayer02}.
Here and further we neglect the magnetic field effect on internal
orbital motion of electrons and holes in the exciton which has been
extensively studied in the past \cite{Elliott60, Gor'kov67, Lerner80,
Lozovik02}. That orbital effect is independent on spin and results in
the exciton diamagnetic shift similar to the blue shift of atomic lines
due to the Langevin diamagnetism.

As we show below, the non-trivial behavior of the equilibrium state of
the condensate in magnetic field is due to the possibility of resonant
scattering of dark to bright excitons and vice versa described by the
last term in the Hamiltonian (\ref{Ham0}). Minimization of $H+H_{Z}$
over $\psi_{m}^*$ yields a relation between the amplitudes of the spin
components
\begin{multline}
 \label{EqnFourAm}
 (V_xA_{-m}^2+V_eA_{l}^2+V_hA_{-l}^2)A_{m}+WA_{-m}A_{l}A_{-l}
 \\ =\left(V_0n-\mu-\frac{1}{2}\omega_m\right)A_{m},
\end{multline}
where $l=\pm 2,\pm 1$ and $\omega_{m}=\pm \omega_{1,2}$ for $m=\pm
1,\pm 2$, respectively. There are solutions to Eqs.\ (\ref{EqnFourAm})
describing FCC, two-component (TCC) and one-component condensate (OCC).
We note that there is no solution with three components---the equation
corresponding to the single empty component cannot be satisfied due to
the $W$-term in (\ref{EqnFourAm}). Physically, it becomes clear if we
remember that each exciton is composed by an electron in $+1/2$ or
$-1/2$ state and a hole in $+3/2$ or $-3/2$ state. As soon as the
electron (or hole) upper spin component is emptied, the transition from
FCC to TCC takes place. If both electrons and holes are fully
spin-polarized we obtain OCC.

To describe the polarization of FCC we introduce auxilliary variables
\begin{equation}
 \label{ParamT}
 t\equiv t_1=\frac{A_{+1}A_{-1}}{A_{+2}A_{-2}}, \quad
 t^{-1}\equiv t_2=\frac{A_{+2}A_{-2}}{A_{+1}A_{-1}},
\end{equation}
that make it possible to write the concentrations, $n_1=A_{+1}^2+A_{-1}^2$ and
$n_2=A_{+2}^2+A_{-2}^2$, and pseudospins of the bright and dark components as
\begin{equation}
 \label{N12}
 n_{1,2}=\frac{Wt_{1,2}+V_x-V_e-V_h}{W(t_1+t_2)+2(V_x-V_e-V_h)}n,
\end{equation}
\begin{equation}
 \label{S12}
 s_{1,2}=\frac{(Wt_{1,2}+V_x)\omega_{1,2}+(V_e-V_h)\omega_{2,1}}%
              {(Wt_1+V_x)(Wt_2+V_x)-(V_e-V_h)^2}n.
\end{equation}
Substitution of these expressions back to definition (\ref{ParamT}) allows to
link $t$ and applied magnetic field by the equation
\begin{equation}
 \label{BofT}
 s_1^2-t^2s_2^2=n_1^2-t^2n_2^2.
\end{equation}
Since $s_{1,2}\propto B$ this equation directly expresses magnetic field as a
function of $t$. Finally, the change of the chemical potential of FCC in
magnetic field is given by
\begin{equation}
 \label{mu4}
 \mu-\mu_0=\frac{n}{4}\frac{W(W+V_e+V_h-V_x)(t-1)^2}{W(t+1)^2-2t(W+V_e+V_h-V_x)}.
\end{equation}

The system (\ref{EqnFourAm}) can be easily solved for TCC and OCC. The
signs of $g$-factors of bright and dark excitons define which
components remain occupied in high magnetic fields. In what follows, we
consider the case $g_{1}\geqslant g_{2}\geqslant 0$, when applied
magnetic field favors the formation of right-circular condensates
\cite{NoteG}. Then, for TCC the amplitudes of components are
\begin{equation}
 \label{TCC}
 A_{+1,2}^2=\frac{1}{2}n+\frac{\omega_{1,2}-\omega_{2,1}}{4V_e}, \quad
 A_{-1,2}^2=0,
\end{equation}
and the chemical potential is
\begin{equation}
 \label{mu2}
 \mu-\mu_0=\frac{n}{4}(W+V_x-V_e+V_h)-\frac{1}{4}(\omega_1+\omega_2).
\end{equation}

In the region of very strong magnetic fields, the condensate becomes
one-component, $A_{+1}=\sqrt{n}$, $A_{-1}=A_{\pm2}=0$. The chemical potential
of OCC is
\begin{equation}
 \label{mu1}
 \mu-\mu_0=\frac{n}{4}(W+V_x+V_e+V_h)-\frac{1}{2}\omega_1.
\end{equation}

\begin{figure}[t]
\includegraphics[width=3.4in]{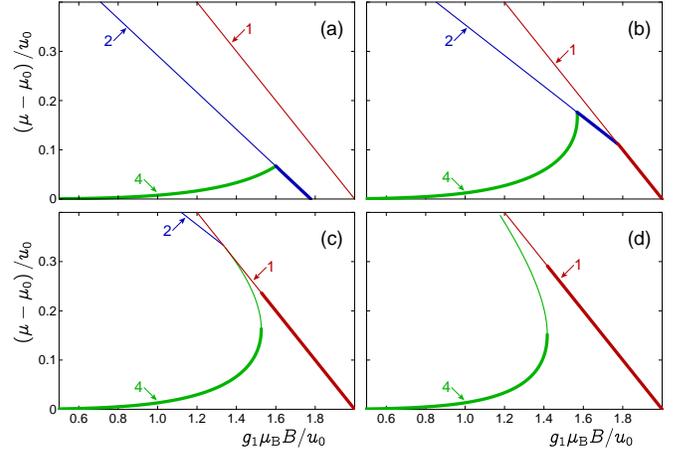}
 \caption{The magnetic-field dependencies of chemical potentials of
 four-component (green lines), two-component (blue lines), and
 one-component (red lines) exciton condensates.
 (the curves are also labeled by the number of components).
 Stable and metastable states of a uniform condensate are indicated by
 bold and thin parts of the curves, respectively.
 The interaction parameters are:
 (a) $W=V_x$, $V_e=V_h=0.5V_x$, $g_2/g_1=0.5$;
 (b) $W=V_x$, $V_e=V_h=0.5V_x$, $g_2/g_1=0.25$;
 (c) $W=V_x$, $V_e=(3/8)V_x$, $V_h=(5/8)V_h$, $g_2/g_1=0.25$;
 (d) $W=3V_x$, $V_e=V_h=0.5V_x$, $g_2/g_1=0.25$.}
\end{figure}

The chemical potential of FCC is independent of magnetic field in the
case of equal $g$-factors $g_1=g_2$, since the bright and dark excitons
are polarized in the same way and $t=1$. However, the behavior of $\mu$
becomes nontrivial if $g_1\ne g_2$. In this case, if $W+V_x>V_e+V_h$,
the chemical potential \emph{increases} with magnetic field as one can
see from (\ref{mu4}). From the experimental point of it leads to a
blue-shift of the emission line, i.e., to an apparent diamagnetic
effect. Unlike the conventional diamagnetism, this effect is only
related to spin interactions and does not depend on the orbital motion
of electrons and holes. It is specific to four-component exciton
condensates and does not take place in two-component exciton-polariton
condensates \cite{Rubo06,Larionov10}, where the chemical potential
remains independent of the magnetic field up to some critical field.

The effect we discuss can be referred to as \emph{pseudo-diamagnetism},
since it is not related to the increase of the total energy of the
system. In fact, the Zeeman energy (\ref{HamZ}) decreases quadratically
with magnetic field, because $s_{1,2}\propto B$ for small $B$ [see
(\ref{S12})]. On the other hand, the chemical potential increases
$\propto B^4$ for small $B$. Clearly, for weak fields this increase can
be neglected compared to the usual, orbital diamagnetism $\propto B^2$.
However, as we show below, the additional pseudo-diamagnetic shift
becomes dominant when the bare Zeeman energy approaches the energy of
exchange interactions in the exciton condensate and it can result in
dramatic changes in the polarization state of the system.

The behavior of chemical potentials (\ref{mu4}), (\ref{mu2}), and (\ref{mu1})
as functions of applied magnetic field is shown in Fig.~2(a-d). For a small
difference of $g$-factors the changes of the ground state of the condensate are
continuous, as it is shown in Fig.~2(a,b). The chemical potential of FCC
slightly increases, and at some magnetic field it reaches the chemical
potential of TCC. At this point the FCC transforms into TCC: the components
$A_{-2}$ and $A_{-1}$ vanish. Subsequently, the TCC is transformed into OCC at
the higher magnetic field, as it is shown in Fig.~2(b). Note also, that FCC can
transform directly into OCC in some range of parameters.

Very interestingly, in a wide range of parameters, namely, for a sufficiently
large amplitude of the mixing term $W$ and strong difference of $g$-factors,
the change of the ground state of the condensate is discontinuous [see
Fig.~2(c,d)]. In these cases, for a given magnetic field there are two
solutions of Eq.~(\ref{BofT}) for parameter $t$ corresponding to different
polarization states of the condensate. The state with a higher value of $\mu$
is metastable. The FCC disappears at $B>B_c$, where $B_c$ is defined by
$d\mu/dB=\infty$. If $B$ reaches $B_c$ from below the FCC with finite
occupation of all components transforms discontinuously into OCC. The chemical
potential jumps by a finite value in this case.

The discontinuous change of the chemical potential is characteristic for a
phase transition of the first-order. In this case one can expect formation of a
mixed state of the system for $B$ close to $B_c$, where FCC and OCC with
different concentrations of excitons coexist, similarly to how it happens in
the case of vapor-liquid phase transition.

In conclusion, we have shown that applied magnetic field suppresses the mixing
of dark and bight excitons and leads to pseudo-diamagnetic increase of the
chemical potential of exciton condensate, provided the dark and bright excitons
possess different $g$-factors. The interplay between spin-dependent
exciton-exciton interactions and Zeeman effect can lead to the first order
transition between four-component and one- or two-component condensates.

We are grateful to B.\ L.\ Altshuler and M.\ M.\ Glazov for valuable
discussions. This work was supported in part by DGAPA-UNAM under Project No.\
IN112310 and by the EU FP7 IRSES Project POLAPHEN.




\begin{thebibliography}{99}
\bibitem{NoteG} The other cases can be reproduced by simple relabeling of the
    components and interaction constants, since Eqs.\ (6) are invariant under
    simultaneous transformations $V_e\leftrightarrow V_h$, $A_{+2}\leftrightarrow
    A_{-2}$, and $g_2\rightarrow -g_2$.
\end{thebibliography}
\end{document}